\documentstyle[emulateapj,apjfonts,epsfig,graphicx]{article}
\slugcomment{}

\begin{document}

\title{Possible new $\gamma$-ray pulsar detections for AGILE and GLAST
missions: \\ The outer gap model look at the Parkes pulsar
Catalog}

\author{Diego F. Torres$^\dag$ \& Sebasti\'an E. Nuza$^\ddag$}
\affil{$^\dag$ Lawrence Livermore National Laboratory, 7000 East
Ave. L$-$413, Livermore, CA 94550, USA } \affil{$^\ddag$
Departamento de Física J.J. Giambiagi, FCEN, UBA, Pabellón 1,
Ciudad Universitaria, 1428 Buenos Aires, Argentina}

\begin{abstract}

The predictive power of the outer gap model of high energy
emission from pulsars is used to analyze the Parkes Multibeam
pulsar survey. We find that most of the radio pulsars of the
Parkes catalog are not $\gamma$-ray emitters according to the
outer gap model. The sample of possible new $\gamma$-ray pulsar
detections for AGILE and GLAST is given. That includes thirteen
new excellent candidates. Four new positional coincidences between
EGRET detections and Parkes pulsars are found, but for which we
discard a physical association. The consequences of applying a new
electron density model in assigning the pulsar distances are
explored. The new model systematically reduce the distances to the
pulsars, corrections can be as large as 90\%, increasing their
fluxes and affecting the detection prospects.
\end{abstract}


\section{Introduction}

The Parkes multibeam pulsar survey (PMPS, now containing 468
objects) is a large-scale survey of a narrow strip of the inner
Galactic plane ($|b|< 5\arcdeg$, $260\arcdeg < l < 50\arcdeg$;
Manchester et al. 2001). PMPS pulsars are generally thought to be
potential counterparts of EGRET unidentified sources. However,
only a handful of them are superposed with EGRET detections, and
even less can be considered a plausible counterpart  (e.g. D'Amico
et al. 2001; Camilo et al. 2001; Torres et al. 2001b). PMPS
pulsars are also thought of as natural candidates for new
$\gamma$-ray detections by AGILE and GLAST missions. Nevertheless,
this is strongly dependent on the high energy emission model
assumed. Predicting which pulsars can (or cannot) be emitting
$\gamma$-rays is essential in determining if these models are
correct.

\section{The outer gap model
 look at the Parkes catalog}

There are basically two kinds of models for high energy emission
from pulsars.
In polar cap models, charged particles are accelerated in
charged-depleted zones near the poles of the pulsar. $\gamma$-rays
are produced through curvature-radiation-induced $\gamma$-B pair
cascades (Harding 1981) or Compton-induced pair cascades (Dermer
\& Sturner 1994). For outer gaps, particle acceleration occurs in
charged-depleted regions in the outer magnetosphere. There are two
outer gap models: thin and thick outer gaps. Here we shall use the
thick outer gap scenario because  it can be applied to all radio
pulsars, including mature ones (this is the case of most  pulsars
in our sample, whose mean characteristic age is 500 kyr).

In the thick outer gap model (Zhang \& Cheng 1997), the size of
the outer gap, $f_s$, is limited by pair production between the
soft thermal X-rays from the stellar surface and the curvature
photons with energy $E_\gamma (f_s)$ emitted by the primary
e$^{\pm}$ accelerated in the gap. The energy of the soft X-ray
photons, $E_X(f_s)$, is determined by the backflow of the gap's
e$^{\pm}$. Therefore $E_X(f_s)$ is also a function of the gap
size. Using $E_X(f_s)E_\gamma (f_s) \sim (mc^2 )^2$ , the size of
the outer gap (the ratio between the outer gap volume and $R_L^3$,
with $R_L$ the light cylinder radius) is: $
 f_s = 5.5 P^{ 26/21} B_{12}^{- 4/7},
$ where $P$ is the pulsar period and $B_{12}$ the magnetic field
in units of $10^{12}$ G (see Zhang \& Cheng 1997; Zhang \& Cheng
1997; 1998a,b for details). It should be noted that $f_s$ is bound
to be less than 1 for this model to make sense. The $\gamma$-ray
luminosity is: $
 L_\gamma \sim 3.6 \times 10^{31} f_s^3 B_{12}^2 P^{-4} {\rm erg} \;{\rm s}^{-1}.
$
The integrated flux on Earth is then given by
 \begin{equation}
F_\gamma^{th} \sim \frac{L_\gamma}{\Delta \Omega d^2 \bar{E_c} }
\sim \frac{7\times 10^{-7}}{d_{kpc}^2\Delta \Omega} \left(
\frac{B_{12} }{P} \right)^{11/28} {\rm ph\;cm}^{-2}\;{\rm s}^{-1}.
 \end{equation}
Here, $\bar {E_c}$ is an average energy of the radiated photons,
$\bar{E_c} \sim 10^{-3} (P/B_{12})^{3/28}$ erg, $\Delta \Omega$ is
the beaming angle, and $d$ the distance to the pulsar. The
theoretical efficiency of the model $\eta_{th}=L_\gamma/\dot E$
results in
$\eta_{th} \sim 83 B_{12}^{-12/7} P^{26/7}.$
These formulae has been compared with data for the observed
$\gamma$-ray pulsars, and for 350 radio pulsars with ages above 1
kyr and known $\gamma$-ray flux upper limits (Zhang \& Cheng
1998a, see also Zhang et al. 2000). Zhang \& Cheng (2002) have
considered the outer gap model in relation with the soft gamma-ray
repeaters, and Anchordoqui et al. (2002) used an outer gap
scenario to study the neutrino production in X-ray binaries.

Tables 1 and 2 show PMPS pulsars for which $f_s < 1$. Columns show
the pulsar name, the period and period derivative, the
characteristic age $\tau=P/2\dot P$, the spin-down luminosity
($\dot E=4\pi^2 I \dot P /P^3$, with $I=10^{45}$\,g\,cm$^2$), the
magnetic field (assuming a dipole model, $\dot E = 2.8 \times
10^{31} B_{12}^2 P^{-4}$ erg s$^{-1}$, with $B_{12}$ being the
magnetic field in units of $10^{12}$ G), the distance (obtained
using the model by Taylor and Cordes 1993), $f_s$, the theoretical
expected efficiency, the $\gamma$-ray luminosity, the flux on
Earth (assuming 1 sr beaming), and the comparison of the predicted
flux with the sensitivity of forthcoming satellites. A y-mark
represents that the flux is above AGILE (GLAST) sensitivity,
considered as 5$\times 10^{-8}$ ($2\times 10^{-9}$) photons
cm$^{-2}$ s$^{-1}$, respectively. It is clear that different
beaming fractions can change the verdict on which pulsar will be
possible to observe. Thus, the detection marks y and n are only
indicative of what can one expect, but are not to be considered
final predictions. On the contrary, PMPS pulsars not contained in
these tables are unable to produce $\gamma$-rays if the outer gap
model is correct, no matter any other possible uncertainty in
distance or beaming angle.

Table 1 contains pulsars for which the efficiency is less than
20\%, as for the observed 3EG $\gamma$-ray pulsars. Table 2 lists
pulsars which have $f_s<1$, but produce much larger values of
theoretical efficiencies. 60\% of these pulsars have $\dot
E_{33}/d^2<0.5$, which has been often used as an indicator for a
low probability of detection (e.g. Thompson et al. 2001). While
these pulsars could still be plausible outer gap $\gamma$-ray
pulsar candidates from a strictly theoretical point of view, they
are not phenomenologically favored.

\begin{table*}
\begin{center}
\caption{Parkes pulsars with $f_s<1$ and $\eta_{th} <0.2$. Those
pulsars having a star present $\dot E > 10^{36}$ erg s$^{-1}$,
$\dot E_{33}/d^2 \gg 0.5$, and very low values of $f_s$ and
$\eta_{th}$. Those with a $^\dag$-symbol are in coincidence with
EGRET sources. }\vspace{0.2cm} {\small
\begin{tabular}{lllllllllllll}
\hline Pulsar J & $P$ & $\dot P$ & $\tau$ & $\dot E$ & $B$ & $d $
& $f_s$ & $\eta_{th} $& $L_\gamma$ &
 $F_{th}^{\Delta \Omega=1}$ & A? & G? \\
& ms & $10^{-15}$ & kyr & $10^{33}$ erg s$^{-1}$ & $10^{12}$ G &
kpc & & &$10^{33}$ erg s$^{-1}$  &10$^{-8}$ cm$^{-2}$ s$^{-1}$ &\\

\noalign{\smallskip} \hline \noalign{\smallskip}

J0834$-$4159                &  121    &  4.4    &  432.8  &  98.6   &  0.9    &  9.7  /1.6  &  0.43 &  0.04 &  10.5     &  1.6   /58.9  &  n  /y&  y  \\
J0855$-$4644$^\star$        &  65     &  7.3    &  141.2  &  1059.3 &  0.8    &  9.9  /3.8  &  0.21 &  $<0.01$ &  12.3  &  1.9   /12.9  &  n  /y&  y  \\
J0901$-$4624                &  442    &  87.5   &  80.1   &  40.0   &  7.3    &  7.4  /2.8  &  0.64 &  0.13 &  13.3     &  3.8   /26.9  &  n  /y&  y  \\
J0940$-$5428$^\star$        &  88     &  32.9   &  42.2   &  1933.9 &  2.0    &  4.2  /2.9  &  0.18 &  $<0.01$ &  14.6  &  13.1  /28.4  &  y  /y&  y  \\
J1015$-$5719$^\dag$         &  140    &  57.4   &  38.7   &  827.5  &  3.3    &  4.8  /5.0  &  0.24 &  0.00 &  14.8     &  10.2  /9.6   &  y  /y&  y  \\
J1016$-$5819                &  88     &  0.7    &  1995.1 &  40.7   &  0.3    &  4.6  /4.7  &  0.54 &  0.08 &  8.4      &  5.3   /5.1   &  y  /y&  y  \\
J1016$-$5857$^\dag$$^\star$ &  107    &  80.6   &  21.1   &  2570.5 &  3.4    &  9.3  /8.0  &  0.17 &  $<0.01$ &  16.1  &  3.1   /4.1   &  n  /n&  y  \\
J1019$-$5749                &  162    &  20.1   &  128.3  &  184.7  &  2.1    &  30.0 /6.9  &  0.37 &  0.02 &  12.5     &  0.2   /3.7   &  n  /n&  y  \\
J1052$-$5954                &  181    &  20.0   &  143.3  &  133.9  &  2.2    &  13.5 /8.5  &  0.41 &  0.03 &  12.3     &  1.0   /2.5   &  n  /n&  y  \\
J1112$-$6103$^\star$        &  65     &  31.5   &  32.7   &  4530.4 &  1.7    &  30.0 /12.2 &  0.13 &  $<0.01$ &  15.2  &  0.3   /1.8   &  n  /n&  y  \\
J1119$-$6127$^\star$        &  408    &  4021.8 &  1.6    &  2342.1 &  47.5   &  30.0 /17.1 &  0.19 &  $<0.01$ &  23.3  &  0.5   /1.5   &  n  /n&  y  \\
J1138$-$6207                &  118    &  12.5   &  149.4  &  303.2  &  1.4    &  24.5 /9.6  &  0.31 &  0.01 &  12.2     &  0.3   /1.9   &  n  /n&  y  \\
J1156$-$5707                &  288    &  26.5   &  172.9  &  43.5   &  3.2    &  20.4 /6.0  &  0.60 &  0.10 &  12.0     &  0.4   /4.6   &  n  /n&  y  \\
J1248$-$6344                &  198    &  16.9   &  185.9  &  85.6   &  2.1    &  23.4 /8.3  &  0.47 &  0.05 &  11.8     &  0.3   /2.3   &  n  /n&  y  \\
J1301$-$6305$^\star$        &  185    &  266.7  &  11     &  1676.0 &  8.2    &  15.8 /7.8  &  0.20 &  $<0.01$ &  17.7  &  1.3   /5.3   &  n  /y&  y  \\
J1327$-$6400                &  281    &  31.2   &  142.7  &  55.7   &  3.5    &  30.0 /15.4 &  0.56 &  0.08 &  12.3     &  0.2   /0.7   &  n  /n&  y  \\
J1406$-$6121                &  213    &  54.7   &  61.8   &  223.2  &  4.0    &  9.1  /8.1  &  0.36 &  0.02 &  13.8     &  2.7   /3.4   &  n  /n&  y  \\
J1412$-$6145$^\dag$         &  315    &  98.7   &  50.7   &  124.3  &  6.5    &  9.3  /7.8  &  0.45 &  0.04 &  14.2     &  2.6   /3.7   &  n  /n&  y  \\
J1413$-$6141$^\dag$         &  286    &  333.4  &  13.6   &  564.9  &  11.4   &  11.0 /10.1 &  0.28 &  0.01 &  17.2     &  2.5   /2.9   &  n  /n&  y  \\
J1420$-$6048$^\dag$$^\star$ &  68     &  83.2   &  13     &  10359.6&  2.8    &  7.6  /5.6  &  0.11 &  $<0.01$ &  17.3  &  5.1   /9.6   &  y  /y&  y  \\
J1452$-$5851                &  387    &  50.7   &  120.9  &  34.6   &  5.2    &  5.6  /4.3  &  0.66 &  0.14 &  12.6     &  6.1   /10.4  &  y  /y&  y  \\
J1509$-$5850                &  89     &  9.2    &  153.7  &  514.9  &  1.1    &  3.7  /2.5  &  0.26 &  $<0.01$ &  12.2  &  13.0  /29.7  &  y  /y&  y  \\
J1514$-$5925                &  149    &  2.9    &  818.3  &  34.5   &  0.8    &  4.3  /3.5  &  0.60 &  0.11 &  9.6      &  6.9   /10.8  &  y  /y&  y  \\
J1524$-$5625$^\star$        &  78     &  39.0   &  31.8   &  3213.2 &  2.0    &  3.8  /2.8  &  0.15 &  $<0.01$ &  15.2  &  17.2  /32.1  &  y  /y&  y  \\
J1531$-$5610                &  84     &  13.7   &  97.2   &  908.7  &  1.3    &  3.0  /2.0  &  0.22 &  $<0.01$ &  13.0  &  21.2  /50.1  &  y  /y&  y  \\
J1538$-$5551                &  105    &  3.2    &  517.3  &  110.4  &  0.7    &  10.3 /7.4  &  0.41 &  0.03 &  10.2     &  1.4   /2.7   &  n  /n&  y  \\
J1541$-$5535                &  296    &  75.0   &  62.5   &  114.4  &  5.5    &  7.5  /5.7  &  0.45 &  0.04 &  13.8     &  3.9   /6.8   &  n  /y&  y  \\
J1543$-$5459                &  377    &  52.0   &  114.9  &  38.3   &  5.2    &  6.3  /4.8  &  0.64 &  0.13 &  12.7     &  4.9   /8.4   &  n  /y&  y  \\
J1548$-$5607                &  171    &  10.7   &  252.4  &  84.9   &  1.6    &  6.9  /4.8  &  0.47 &  0.05 &  11.3     &  3.5   /7.3   &  n  /y&  y  \\
J1601$-$5335                &  288    &  62.4   &  73.3   &  102.6  &  5.0    &  4.0  /4.5  &  0.47 &  0.05 &  13.5     &  13.1  /10.5  &  y  /y&  y  \\
J1626$-$4807                &  294    &  17.5   &  266.7  &  27.2   &  2.7    &  10.2 /8.9  &  0.69 &  0.16 &  11.2     &  1.6   /2.1   &  n  /n&  y  \\
J1632$-$4757                &  229    &  15.1   &  240.4  &  49.8   &  2.2    &  7.0  /6.3  &  0.56 &  0.09 &  11.4     &  3.4   /4.2   &  n  /n&  y  \\
J1632$-$4818                &  813    &  650.4  &  19.8   &  47.7   &  27.0   &  8.5  /7.7  &  0.64 &  0.13 &  16.3     &  3.8   /4.6   &  n  /n&  y  \\
J1637$-$4642$^\dag$         &  154    &  59.2   &  41.2   &  639.6  &  3.5    &  5.7  /5.1  &  0.26 &  $<0.01$ &  14.7  &  7.2   /9.2   &  y  /y&  y  \\
J1638$-$4608                &  278    &  51.5   &  85.6   &  94.5   &  4.4    &  5.8  /5.1  &  0.48 &  0.05 &  13.2     &  6.1   /8.0   &  y  /y&  y  \\
J1648$-$4611                &  165    &  23.7   &  110.1  &  208.9  &  2.3    &  5.7  /5.0  &  0.36 &  0.02 &  12.7     &  6.1   /7.9   &  y  /y&  y  \\
J1702$-$4128                &  182    &  52.3   &  55.2   &  342.0  &  3.6    &  5.1  /4.7  &  0.32 &  0.01 &  14.1     &  8.5   /10.2  &  y  /y&  y  \\
J1702$-$4310                &  241    &  223.8  &  17     &  634.9  &  8.6    &  5.4  /5.1  &  0.27 &  0.01 &  16.6     &  9.6   /10.9  &  y  /y&  y  \\
J1705$-$3950                &  319    &  60.6   &  83.4   &  73.7   &  5.2    &  3.8  /3.2  &  0.52 &  0.07 &  13.3     &  14.0  /20.3  &  y  /y&  y  \\
J1715$-$3903$^\dag$         &  278    &  37.7   &  117.2  &  68.9   &  3.8    &  4.7  /4.1  &  0.52 &  0.07 &  12.6     &  8.7   /11.6  &  y  /y&  y  \\
J1718$-$3825$^\star$        &  75     &  13.2   &  89.5   &  1253.7 &  1.2    &  4.2  /3.6  &  0.20 &  $<0.01$ &  13.1  &  11.5  /15.9  &  y  /y&  y  \\
J1723$-$3659                &  203    &  8.0    &  401.4  &  37.9   &  1.5    &  4.2  /3.5  &  0.60 &  0.11 &  10.6     &  8.4   /12.5  &  y  /y&  y  \\
J1726$-$3530                &  1110   &  1216.8 &  14.5   &  35.1   &  43.1   &  9.9  /8.4  &  0.72 &  0.19 &  17.0     &  3.0   /4.2   &  n  /n&  y  \\
J1734$-$3333                &  1169   &  2279.0 &  8.1    &  56.3   &  60.5   &  7.4  /6.4  &  0.64 &  0.13 &  18.5     &  6.0   /8.0   &  y  /y&  y  \\
J1735$-$3258                &  351    &  26.1   &  213.4  &  23.8   &  3.5    &  11.1 /9.5  &  0.72 &  0.19 &  11.6     &  1.4   /1.9   &  n  /n&  y  \\
J1737$-$3137                &  450    &  138.8  &  51.5   &  59.9   &  9.3    &  5.8  /5.4  &  0.57 &  0.09 &  14.2     &  6.7   /7.9   &  y  /y&  y  \\
J1738$-$2955                &  443    &  81.9   &  85.9   &  37.1   &  7.1    &  3.9  /3.4  &  0.65 &  0.14 &  13.2     &  13.6  /17.9  &  y  /y&  y  \\
J1739$-$3023                &  114    &  11.4   &  159    &  300.9  &  1.3    &  3.4  /2.9  &  0.31 &  0.01 &  12.1     &  15.8  /21.8  &  y  /y&  y  \\
J1743$-$3153                &  193    &  10.6   &  289.7  &  57.9   &  1.7    &  8.0  /6.5  &  0.53 &  0.07 &  11.1     &  2.5   /3.8   &  n  /n&  y  \\
J1806$-$2124                &  482    &  118.9  &  64.2   &  42.0   &  8.9    &  10.0 /9.8  &  0.63 &  0.13 &  13.8     &  2.2   /2.2   &  n  /n&  y  \\
J1809$-$1917$^\star$        &  83     &  25.5   &  51.4   &  1779.8 &  1.7    &  3.7  /3.5  &  0.18 &  $<0.01$ &  14.2  &  16.7  /18.7  &  y  /y&  y  \\
J1828$-$1101$^\star$        &  72     &  14.8   &  77.1   &  1563.0 &  1.2    &  7.2  /6.6  &  0.18 &  $<0.01$ &  13.4  &  4.0   /4.8   &  n  /n&  y  \\
J1837$-$0604$^\dag$$^\star$ &  96     &  45.2   &  33.8   &  1998.6 &  2.4    &  6.1  /6.4  &  0.18 &  $<0.01$ &  15.1  &  6.5   /6.0   &  y  /y&  y  \\
J1838$-$0453$^\dag$         &  381    &  115.7  &  52.2   &  82.7   &  7.8    &  8.2  /8.1  &  0.51 &  0.06 &  14.2     &  3.3   /3.4   &  n  /n&  y  \\
J1839$-$0321                &  239    &  12.5   &  302.4  &  36.3   &  2.0    &  6.8  /7.2  &  0.62 &  0.12 &  11.0     &  3.5   /3.1   &  n  /n&  y  \\
J1843$-$0355                &  132    &  1.0    &  2017.1 &  17.7   &  0.4    &  8.8  /8.8  &  0.72 &  0.18 &  8.4      &  1.4   /1.4   &  n  /n&  y  \\
J1853$+$0056                &  276    &  21.4   &  204.3  &  40.3   &  2.8    &  3.8  /5.1  &  0.61 &  0.11 &  11.7     &  12.0  /6.7   &  y  /y&  y  \\
J1904$+$0800                &  263    &  17.3   &  241.1  &  37.4   &  2.5    &  9.2  /8.6  &  0.62 &  0.12 &  11.4     &  2.0   /2.3   &  n  /n&  y  \\
J1907$+$0345                &  240    &  8.2    &  463.1  &  23.4   &  1.6    &  8.7  /7.1  &  0.70 &  0.17 &  10.4     &  2.0   /3.0   &  n  /n&  y  \\
J1908$+$0909                &  337    &  34.9   &  153    &  36.1   &  4.0    &  8.8  /8.9  &  0.64 &  0.13 &  12.2     &  2.4   /2.3   &  n  /n&  y  \\
J1909$+$0912                &  223    &  35.8   &  98.7   &  127.6  &  3.3    &  8.2  /8.2  &  0.43 &  0.04 &  12.9     &  3.0   /3.0   &  n  /n&  y  \\
J1913$+$0832                &  134    &  4.6    &  466.2  &  74.3   &  0.9    &  7.7  /7.8  &  0.48 &  0.05 &  10.4     &  2.5   /2.4   &  n  /n&  y  \\
J1913$+$1011$^\star$        &  36     &  3.4    &  169.1  &  2871.4 &  0.4    &  4.4  /4.7  &  0.14 &  $<0.01$ &  12.0  &  9.1   /8.2   &  y  /y&  y  \\

\noalign{\smallskip} \hline \noalign{\smallskip}
\end{tabular}
\mbox{}}
\end{center}
\end{table*}

In the thick outer gap model given by Zhang \& Cheng (1997), the
effects of magnetic inclination angle ($\alpha$) are not taken
into account (Zhang \& Cheng 2002). Under these effects, $f_s$
transforms into $f_s \times c(\alpha)$ where $c(\alpha)=$1, 0.9,
0.83, 0.70, 0.57 for inclination angles of 45, 55, 65, and 75
degrees (Cheng 2002). However,  although $f_s$ will vary with
$\alpha$, its value at $R_L/2$, where $R_L$ is the radius of light
cylinder (see Cheng, Ruderman \& Zhang, 2000) is a sensible
approximation. In addition, most of the Parkes pulsars having $f_s
>1$ have indeed $f_s \gg 1$, and so the $\alpha$-dependence is
unimportant here.

\begin{table*}
\begin{center}
\caption{Additional Parkes pulsars with $f_s<1$ but $\eta_{th}
>0.2$. }\vspace{0.2cm} {\small
\begin{tabular}{lllllllllllll}
\hline Pulsar J & $P$ & $\dot P$ & $\tau$ & $\dot E$ & $B$ & $d $
& $f_s$ & $\eta_{th} $& $L_\gamma$ &
 $F_{th}^{\Delta \Omega=1}$ & A? & G? \\
& ms & $10^{-15}$ & kyr & $10^{33}$ erg s$^{-1}$ & $10^{12}$ G &
kpc & & &$10^{33}$ erg s$^{-1}$  &10$^{-8}$ cm$^{-2}$ s$^{-1}$ &\\

\noalign{\smallskip} \hline \noalign{\smallskip}

J0954$-$5430        &  473    &  43.9   &  170.7  &  16.4   &  5.3    &  6.2    /3.9  &  0.83 &  0.29 &  12.0     &  4.7    /11.8   &  n/y  &  y  \\
J0957$-$5432        &  204    &  1.9    &  1657.6 &  9.1    &  0.7    &  7.0    /4.3  &  0.91 &  0.37 &  8.7      &  2.4    /6.3    &  n/y  &  y  \\
J1043$-$6116        &  289    &  10.4   &  439.8  &  17.1   &  2.0    &  18.1   /9.4  &  0.78 &  0.24 &  10.5     &  0.4    /1.4    &  n/n  &  y  \\
J1115$-$6052        &  260    &  7.2    &  569.3  &  16.3   &  1.6    &  6.7    /4.1  &  0.79 &  0.24 &  10.1     &  3.1    /8.4    &  n/y  &  y  \\
J1216$-$6223        &  374    &  16.8   &  352.6  &  12.7   &  2.9    &  30.0   /16.5 &  0.87 &  0.33 &  10.8     &  0.2    /0.6    &  n/n  &  y  \\
J1305$-$6203        &  428    &  32.1   &  211    &  16.2   &  4.3    &  24.1   /8.5  &  0.83 &  0.28 &  11.6     &  0.3    /2.4    &  n/n  &  y  \\
J1349$-$6130        &  259    &  5.1    &  802.4  &  11.6   &  1.4    &  5.8    /4.9  &  0.87 &  0.32 &  9.6      &  4.0    /5.6    &  n/y  &  y  \\
J1452$-$6036        &  155    &  1.4    &  1694.9 &  15.4   &  0.6    &  9.4    /5.8  &  0.76 &  0.22 &  8.6      &  1.3    /3.4    &  n/n  &  y  \\
J1515$-$5720        &  287    &  6.1    &  745.3  &  10.2   &  1.5    &  10.2   /6.6  &  0.91 &  0.37 &  9.7      &  1.3    /3.1    &  n/n  &  y  \\
J1530$-$5327        &  279    &  4.7    &  944.4  &  8.5    &  1.3    &  1.4    /1.2  &  0.95 &  0.43 &  9.4      &  60.8   /90.0   &  y/y  &  y  \\
J1618$-$4723        &  204    &  2.0    &  1619.6 &  9.3    &  0.7    &  3.4    /3.0  &  0.90 &  0.37 &  8.7      &  9.9    /13.0   &  y/y  &  y  \\
J1649$-$4653        &  557    &  49.7   &  177.6  &  11.4   &  6.2    &  5.1    /4.6  &  0.94 &  0.41 &  11.9     &  6.8    /8.5    &  y/y  &  y  \\
J1649$-$4729        &  298    &  6.5    &  720.6  &  9.8    &  1.6    &  12.7   /8.0  &  0.92 &  0.39 &  9.7      &  0.8    /2.0    &  n/n  &  y  \\
J1711$-$4322        &  103    &  0.2    &  7747.6 &  7.7    &  0.2    &  4.1    /3.8  &  0.89 &  0.36 &  6.9      &  4.9    /5.9    &  n/y  &  y  \\
J1812$-$1910        &  431    &  37.7   &  181.1  &  18.6   &  4.7    &  11.5   /11.2 &  0.79 &  0.25 &  11.9     &  1.3    /1.3    &  n/n  &  y  \\
J1835$-$1020        &  302    &  5.9    &  810.2  &  8.4    &  1.6    &  2.5    /2.3  &  0.96 &  0.45 &  9.6      &  20.2   /25.2   &  y/y  &  y  \\
J1837$-$0559        &  201    &  3.3    &  964.6  &  16.1   &  1.0    &  5.0    /5.4  &  0.77 &  0.23 &  9.3      &  5.1    /4.3    &  y/n  &  y  \\
J1842$-$0905        &  345    &  10.5   &  520.7  &  10.1   &  2.2    &  7.4    /5.9  &  0.93 &  0.40 &  10.2     &  2.7    /4.2    &  n/n  &  y  \\
J1845$-$0743        &  105    &  0.4    &  4527.9 &  12.6   &  0.2    &  5.9    /5.2  &  0.77 &  0.23 &  7.5      &  2.7    /3.4    &  n/n  &  y  \\
J1853$+$0545        &  126    &  0.6    &  3279.4 &  11.9   &  0.3    &  4.7    /5.1  &  0.80 &  0.26 &  7.9      &  4.5    /3.8    &  n/n  &  y  \\
J1908$+$0839        &  185    &  2.4    &  1231.7 &  14.8   &  0.8    &  9.4    /9.2  &  0.78 &  0.24 &  9.0      &  1.4    /1.4    &  n/n  &  y  \\

\noalign{\smallskip} \hline \noalign{\smallskip}
\end{tabular}
\mbox{}}
\end{center}
\end{table*}


\section{Pulsar distances}

\begin{figure*}
\centering
\includegraphics[width=0.3\textwidth,height=7cm]{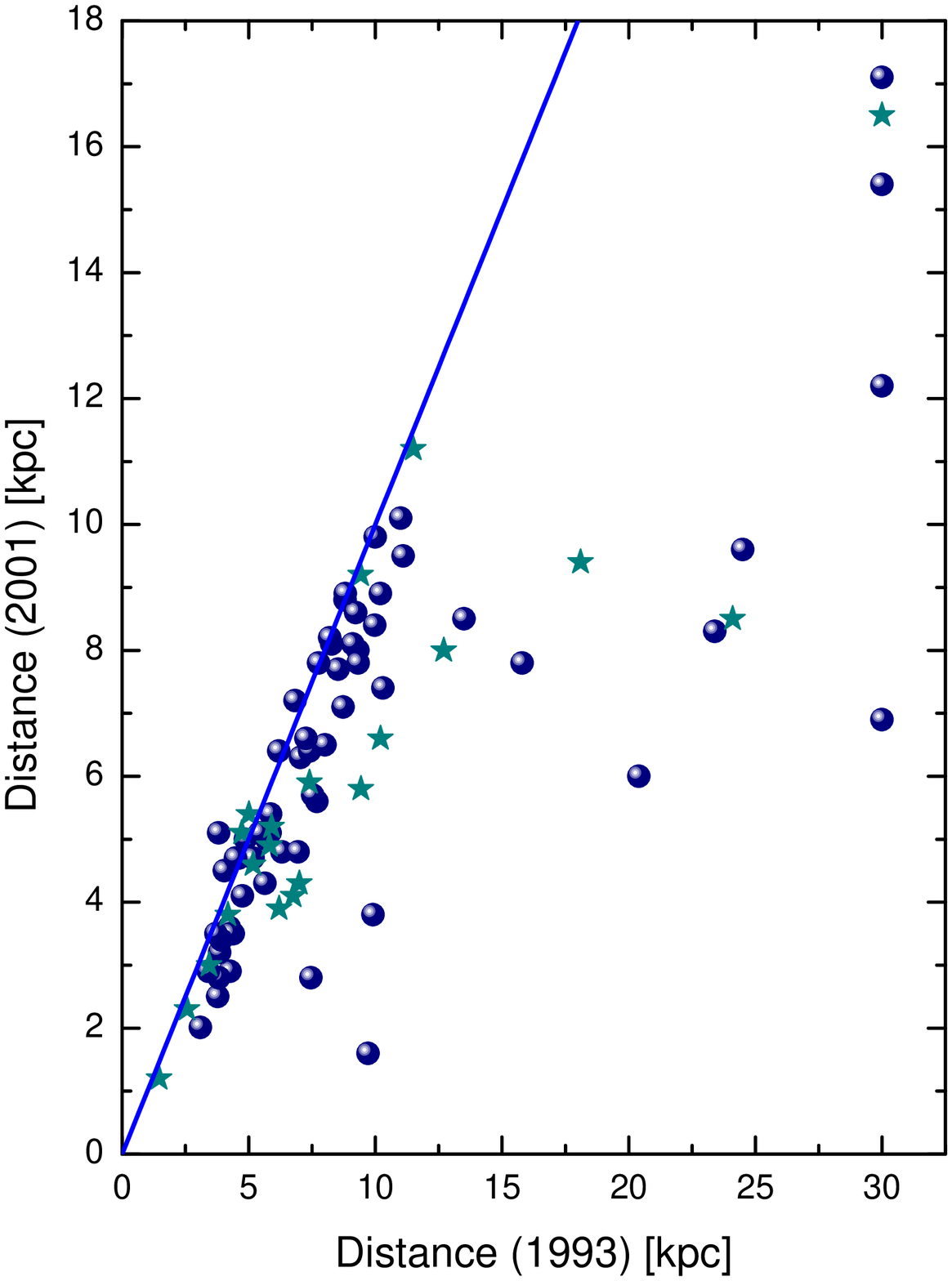}\hspace{2cm}
\includegraphics[width=0.3\textwidth,height=7cm]{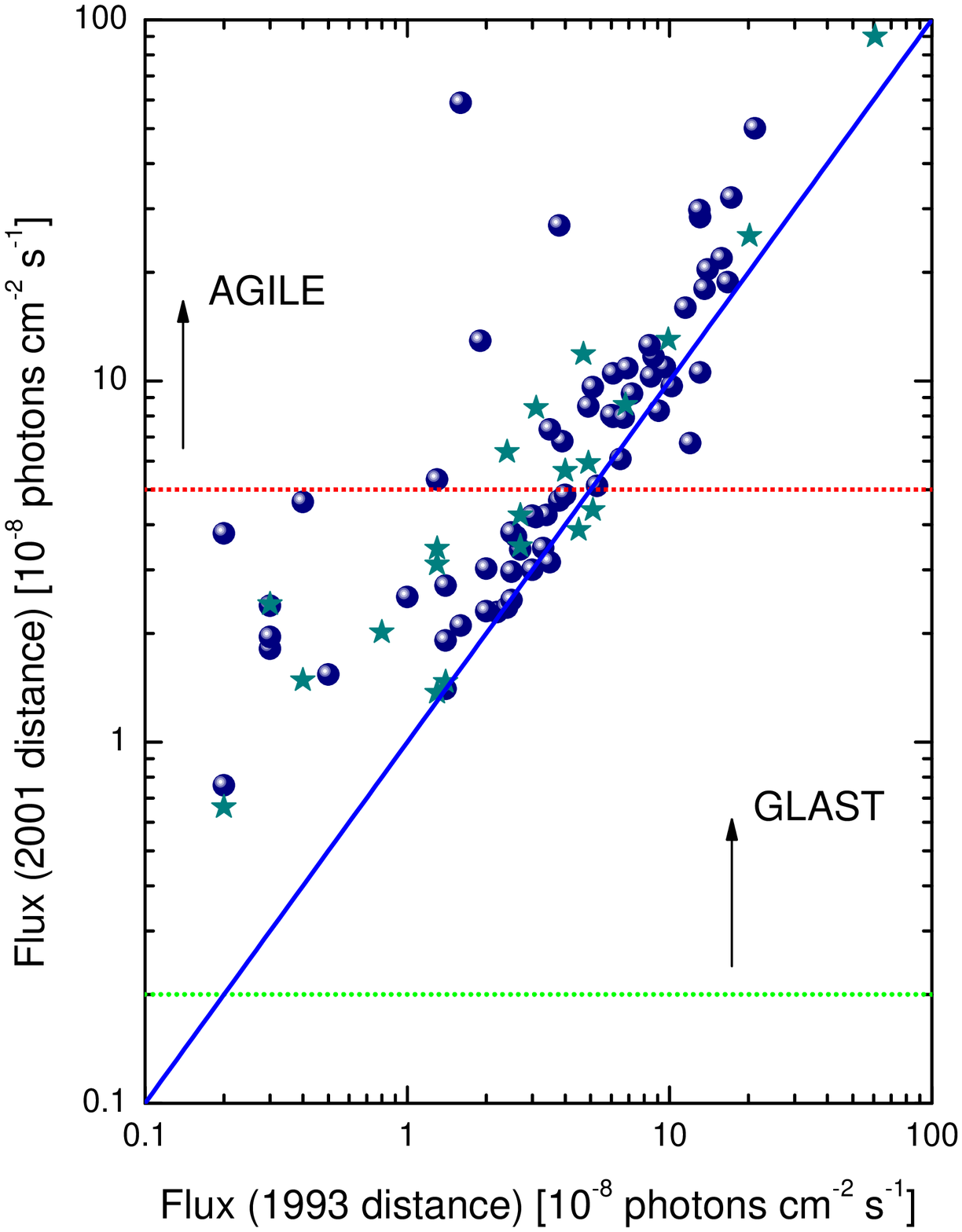}
\caption{Pulsar distances computed using NE2001 and implied
variation in fluxes as compared with the 1993 model results. The
solid line is, in both plots, the curve $y=x$. Circles represent
pulsars in Table 1 ($f_s<1, \eta_{\rm th}<0.2$); stars represent
pulsars in Table 2 ($f_s<1, \eta_{\rm th}>0.2$).}
\end{figure*}

The distances given in the PMPS were obtained applying the model
for the Galactic distribution of electrons given by Taylor and
Cordes (1993). Recently, Cordes and Lazio (2002) presented an
improved model for this distribution. We have used Cordes and
Lazio's (2002) NE2001 code to compute these new distances using
the dispersion measures given in the PMPS. Fig.~1a shows how they
deviate from the previous estimations for all pulsars contained in
Tables 1 and 2. Generally, new inferred values are smaller, what
implies bigger fluxes, see Fig.~1b. For some particular cases
-especially for those pulsars having the largest inferred 1993
distances- differences are notorious. Tables 1 and 2 give, as
well, the newly computed distances and fluxes (separated by a
/-symbol). These values affect the detection criterion for the
AGILE mission. Using the newest electron model, there are 14\%
more $f_s<1$ pulsars with fluxes above the AGILE threshold. Note,
however, that changing the distances to these new values does not
affect the sample of possible outer gap pulsars: The latter is
constructed using only intrinsic pulsar parameters.

\section{Positional coincidences }

There are four EGRET sources now found to be spatially coincident
with newly discovered PMPS pulsars. Table 3 shows these new
positional coincidences, obtained using FORREST (Sigl et al.
2001). In all these cases, the efficiencies required to produce
the $\gamma$-ray sources are $\eta \gg 1000\%$ (even for the
reduced NE2001 distances), making the potential associations
unphysical. None of the pulsars contained in Table 3 has $f_s<1$.

\begin{table*}
\begin{center}
\caption{New Parkes pulsars spatially coincident with EGRET
sources. }\vspace{0.2cm}{\small
\begin{tabular}{llllllllllllllll}
\hline EGRET & Pulsar J & $P$ & $\dot P$ &  $\dot E$ & $\dot
E/d^2$ [1993] & $\dot
E/d^2$ [2001] \\
& & s & $10^{-15}$  & ergs s$^{-1}$ & ergs s$^{-1}$ kpc$^{-1}$ & ergs s$^{-1}$ kpc$^{-1}$ \\
\noalign{\smallskip} \hline \noalign{\smallskip}

1824$-$1514  &   1825$-$1526   & 1.62  & 4.2 & $3.8 \times
10^{31}$ &
$4.4 \times 10^{29}$ &  $5.8 \times 10^{29}$ \\

           &   1826$-$1526   & 0.38  & 1.0 & $7.6 \times 10^{32}$ & $6.4
\times 10^{30}$ &$5.1 \times 10^{31}$ \\

1903$+$0550  &   1905$+$0616   & 0.98  & 135.21 & $5.5 \times
10^{33}$ &
$1.9 \times 10^{32}$ & $1.7 \times 10^{32}$ \\

1638$-$5155  &   1638$-$5226   & 0.34  & 2.65& $2.6 \times
10^{33}$ &
$9.8 \times 10^{31}$ & $2.3 \times 10^{32}$ \\

1704$-$4732 & 1707$-$4729 & 0.26 & 1.5 & $3.3 \times 10^{33}$ &
$2.9 \times 10^{31}$ & $8.5 \times 10^{31}$ \\

\noalign{\smallskip} \hline \noalign{\smallskip}
\end{tabular}

 \label{new$-$p}
\mbox{}\\
}
\end{center}
\end{table*}

The possibly variable source (Torres et al. 2001) 3EG J1824$-$1514
has been proposed to be result of inverse Compton emission from a
microquasar (Paredes et al. 2000). Although there are (considering
all catalogs) three pulsars coincident with this source none pose
a challenge for the microquasar interpretation. There are three
pulsars in the 3EG J1903$+$0550 error box, but none is energetic
enough as to produce it. 3EG J1903$+$0550 is also coincident with
SNR G39.2$-$0.3 (Romero et al. 1999), which is in turn co-spatial
with a giant molecular cloud. Torres et al. (2002) have shown that
the interaction between the molecular cloud and the SNR could
produce most of the flux observed by EGRET, similar to the case of
G347.3-0.5 (Butt et al. 2001).

\section{Concluding Remarks}

In the framework of the outer gap model of $\gamma$-ray emission
from pulsars, we have theoretically computed the $\gamma$-ray
luminosity, the theoretical efficiency, and other parameters for
all radio pulsars listed in the current version of the PMPS. Most
pulsars (82\% out of 468 reported in the latest release) {\it are
not} $\gamma$-ray pulsars if the outer gap model is correct.
Should AGILE or GLAST detect $\gamma$-ray emission coming from
PMPS pulsars not contained in Tables 1 and 2, a different high
energy emission should be operative. This is a definite prediction
valid independently of our ignorance of the beaming angle or the
distance. Tables 1 and 2 then show PMPS pulsars which, in the
framework of the outer gap model, can be $\gamma$-ray detections
for AGILE and GLAST. 13 of these pulsars are excellent candidates,
having large values of $\dot E$ and $\dot E_{33}/d^2$, and low
values of $f_s$ and $\eta_{th}$. The new model for the electron
density in the galaxy (Cordes and Lazio 2002) reduce the pulsar
distances reported in the PMPS, enlarging the predicted fluxes.
The use of this new model is essential in determining the expected
detection; corrections can be as large as 90\%. Finally, four new
PMPS pulsars are coinciding with previously detected EGRET
sources. None of them qualify as a possible counterpart for their
respective spatially coincident $\gamma$-ray source, nor they can
emit $\gamma$-rays if the outer gap model is correct.


The work of D.F.T. was performed under the auspices of the U.S.
D.O.E. (NNSA), by University of California Lawrence Livermore
National Laboratory under contract No. W-7405-Eng-48. D.F.T. is
LLNL's Lawrence Fellow in Astrophysics.  He thanks F. Camilo, G.E.
Romero, C. Mauche, and K.S. Cheng for useful comments.


\end{document}